\documentstyle[12pt]{article}
\input{psfig.sty}
\title{\bf Thermodynamics of a free $SU_{q}(2)$ fermionic system}
\author {\\ \\ \\ Marcelo R. Ubriaco\thanks{E-mail:m\_ubriaco@upr1.upr.clu.edu}\\
{\em Laboratory of Theoretical Physics}\\
{\em Department of Physics}\\
{\em  University of Puerto Rico}\\
{\em P. O. Box 23343, R\'{\i}o Piedras}\\
{\em PR 00931-3343, USA}}
\date{}
\begin{document}
\vspace{0.3in}
\maketitle
\vspace{0.15in}
\begin{abstract}
  
We calculate the partition function, 
average occupation number and internal energy  for a $SU_q(2)$
fermionic system and  compare this model
at $T=0$  with the ordinary fermionic, $q=1$, case.  At low 
temperatures and $q\gg 1$
we find the chemical potential $\mu$ 
to have the same temperature dependence than 
the Fermi case. For $q\ll 1$, 
 the function $\mu(T)$  has in addition a  linear dependence
on $T$.   

\end{abstract}
\baselineskip20pt
\newpage
The role of quantum groups and quantum algebras
 in physics has its origins in the theory
of integrable models and the quantum inverse scattering method
\cite{Jimbo}.  During the past few years a great deal
of interest has been focused on the
relevance that 
quantum groups may play in other fields of physics. Two 
of the most distinct features of quantum group theory to search for
 new
physical applications are:1)
its relation with noncommutative geometry \cite{Wo,Manin}, 
and 2) the fact
that quantum groups  can be seen as 
generalizations (or deformations) of Lie algebras. The first
one led to a concrete  formulation of covariant
 non-commutative differential calculus \cite{WZ} and  
one-parameter quantum group  deformations of bosonic and fermionic
phase spaces \cite{Z}. The second feature
motivated the study of 'deformed' physical systems \cite{U1} in which the
theory becomes the standard one as the deformation parameter
$q\rightarrow 1$. However, the word 'deformation' acquired different
meanings in the literature, not always
 related to the concept of quantum groups.  One of the most typical
examples is the  case of the so called $q$-oscillators, which satisfy
an algebra in terms of deformed commutation relations
rather than an algebra covariant under quantum group transformations.
A recent application of the $q$-fermionic algebra  \cite{NG}
to the Lipkin model can be found in Reference \cite{AEGPL},
and some approaches dealing  with the thermodynamics of $q$-bosons 
are given
in References \cite{MD,S,J,VZ,MO}.

In this article our main purpose
is to study some of the consequences of
considering a "free" quantum group fermionic system. 
We first
briefly discuss  the group $SU_{q}(2)$,
and then give the $SU_{q}(N)$-covariant oscillator algebra 
corresponding
in the $q=1$ limit to a fermionic algebra.
In order to understand how
 $SU_{q}(N)$ oscillators act on the vacuum state we then
 obtain a 
representation of them in terms of
fermionic operators.
For simplicity,
we consider a hamiltonian involving two quantum group flavors, which
in terms of fermionic fields it becomes an interacting fermionic
system.
We then calculate
the partition function, occupation numbers and average energy.
  At $T=0$  the average energy
per particle $\frac{U}{<M>}$ is independent of $q$ and
equal to the Fermi case.  The occupation number as
a function of the energy shows a larger deviation from
usual fermions for the case $q\ll 1$ than $q\gg 1$.  Furthermore, a calculation 
 at low temperatures show that for $q\gg 1$ the chemical potential
$\mu$ is almost identical to the standard one, but for $q\ll 1$
the dependence of $\mu$ on $T$ is radically different than the
$q=1$ case. Therefore, the consequences of
considering quantum group fermionic fields in a model
becomes much more significant when the deformation parameter $q$ is
closer to zero  than to infinity. 

We will denote as  $\Psi_{i}$, $i=1,...,N$,
the quantum group  fermionic operators.
The two dimensional representation of the quantum group  $SU_{q}(2)$ 
\cite{T,VWZ} is given by a matrix
\begin{equation}
T=\left(\begin{array}{cc} a & b \\ c & d\end{array}\right),
\end{equation}
where the matrix coefficients $(a,b,c,d)$ generate the algebra
\begin{eqnarray}
ab=q^{-1}ba  & , & ac=q^{-1}ca \nonumber \\
bc=cb & , & dc=qcd  \nonumber \\
db=qbd & , &  da-ad=(q-q^{-1})bc  \nonumber \\
& & det_{q}T\equiv ad-q^{-1}bc=1  .
\end{eqnarray}
Requiring  $T$ to be unitary leads to the adjoint matrix 
$\overline{T}$ given by
\begin{equation}
\overline{T}=\left(\begin{array}{cc} d & -qb \\ -q^{-1}c & a\end{array}\right),
\end{equation}
where the parameter $q$ must be a real number.  Hereafter, 
we will consider $0\leq q<\infty$ .  Now, given the linear
transformation $\Psi'_{i}=\sum_{j=1}^{2}T_{ij}\Psi_{j}$; 
the  $SU_{q}(2)$-covariant algebra is given
by the following relations \cite{U2}
\begin{equation}
\{\Psi_{2},\overline{\Psi}_{2}\}=1
\end{equation}
\begin{equation}
\{\Psi_{1},\overline{\Psi}_{1}\}=1 - (1-q^{-2})\overline{\Psi}_{2}\Psi_{2}
\end{equation}
\begin{equation} 
\Psi_{1}\Psi_{2}=-q \Psi_{2}\Psi_{1}
\end{equation}
\begin{equation} 
\overline{\Psi}_{1}\Psi_{2}=-q \Psi_{2}\overline{\Psi}_{1}
\end{equation}
\begin{equation}
\{\Psi_{1},\Psi_{1}\}=0=\{\Psi_{2},\Psi_{2}\} \label{0},
\end{equation}
which for $q=1$ become a $SU(2)$ covariant  fermionic algebra.
It is clear from Eq. (\ref{0}) that occupation numbers 
for $SU_{q}(N)$ fermionic states are restricted to
$m=0,1$.
For arbitrary $N$ these relations are written in compact form
as follows
\begin{equation}
\Psi_{j}\overline{\Psi}_{i}=\delta_{ij}
- q^{-1}
\hat{{\cal R}}_{ikjl}
\overline{\Psi}_{l}\Psi_{k}  \label{a}
\end{equation}
\begin{equation}
\Psi_{l}\Psi_{k}=- q \hat{{\cal R}}_{ijkl}
\Psi_{j}\Psi_{i}. \label{b}
\end{equation}
 The matrix
$\hat{{\cal R}}_{ijkl}$  is related to 
 the $R$-matrix $R_{ijkl}$
of $\hat{A}_{N-1}^{q}$ by the relation
\begin{equation}
 \hat{{\cal R}}_{ijkl}=R_{jikl},
\end{equation}
with \cite{WZ}
\begin{equation}
\hat{{\cal R}}_{ijkl}=\delta_{jk}\delta_{il} (1+(q-1)\delta_{ij})
+(q-q^{-1})\delta_{ik}\delta_{jl}\theta(j-i),
\end{equation}
where $\theta(j-i)=1$ for $j>i$ and zero otherwise. 

A representation of $SU_{q}(N)$-fermions in terms of fermionic operators $\psi_{i}$
and $\psi_{j}^{\dagger}$ according to Equations (\ref{a})and (\ref{b}) is given by
\begin{equation}
\Psi_{m}=\psi_{m}\prod_{l=m+1}^{N}\left(1+(q^{-1}-1) M_{l}\right)\label{psi},
\end{equation}
\begin{equation}
\overline{\Psi}_{m}=\psi_{m}^{\dagger}\prod_{l=m+1}^{N}
\left(1+(q^{-1}-1) M_{l}\right)\label{psi1},
\end{equation}
where $M_{l}=\psi_{l}^{\dagger}\psi_{l}$ and $\{\psi_{i},\psi_{j}^{\dagger}\}=
\delta_{ij}$.

We consider the simplest $SU_{q}(2)$ invariant model, which corresponds
to the hamiltonian with two quantum flavors
\begin{equation}
{\cal H}=\sum_{\kappa}^{}\varepsilon_{\kappa}(\overline{\Psi}_{\kappa,1}{\Psi}_{\kappa,1}
+\overline{\Psi}_{\kappa,2}{\Psi}_{\kappa,2}),\label{h}
\end{equation}
with the quantum group fields satisfying $\{\overline{\Psi}_{\kappa,i},
\Psi_{\kappa',j}\}=0$ for $\kappa\neq\kappa'$.
After replacing Eqs.(\ref {psi}) and (\ref{psi1}) into this equation 
the hamiltonian becomes a function of the fermion number operator $M_{\kappa,i}
=\overline{\psi}_{\kappa,i}\psi_{\kappa,i}$ as follows
\begin{equation}
{\cal H}=\sum_{\kappa}^{}\varepsilon_{\kappa}(M_{\kappa,1}+
M_{\kappa,2}+(q^{-2}-1) M_{\kappa,1}M_{\kappa,2}).
\end{equation}
Therefore,  in terms of ordinary fermions the $SU_q(2)$-invariant 
hamiltonian in Equation (\ref {h}) becomes an interacting
$SU(2)$ invariant system with the coupling as a 
function of the parameter $q$.

The grand partition function ${\cal Z}$ is  obtained from the equation
\begin{eqnarray}
{\cal Z}&=&\prod_{\kappa}\sum_{m_{\kappa,1}=0}^{1}
\sum_{m_{\kappa,2}=0}^{1}e^{-\beta \varepsilon_{\kappa}[m_{\kappa,1}
(1+(q^{-2}-1)m_{\kappa,2})+m_{\kappa,2}]}
e^{\beta\mu(m_{\kappa,1}+m_{\kappa,2})} \nonumber \\
&=&\prod_{\kappa}(1+2e^{-\beta (\varepsilon_{\kappa}-\mu)}+
e^{-\beta (\varepsilon_{\kappa}(q^{-2}+1)-2\mu)}).\label{z}
\end{eqnarray}
Equation (\ref{z}) clearly shows that for $q\neq 1$ the partition
function does not factorize in terms of the
internal degrees of freedom. Since the hamiltonian 
is invariant under $M_{\kappa,1}\leftrightarrow M_{\kappa,2}$
interchange, the average number of
particles with energy $\varepsilon_{\kappa}$ is independent of the flavor
index. From the grand potential $\Omega\equiv -\frac{1}{\beta}
ln{\cal Z}$ the average number of particles can be obtained
from the equation
\begin{eqnarray}
<M>&=&-(\frac{\partial\Omega}{\partial\mu})_{T,V} \nonumber \\
&=&2\sum_{\kappa}\frac{e^{-\beta(\varepsilon_{\kappa}-\mu)}(1+e^{-\beta(\varepsilon_{\kappa}q^{-2}
-\mu)})}{1+2e^{-\beta (\varepsilon_{\kappa}-\mu)}+
e^{-\beta (\varepsilon_{\kappa}(q^{-2}+1)-2\mu)}}\nonumber \\
&=&\sum_{\kappa}(<m_{\kappa,1}>+<m_{\kappa,2}>).\label{M}
\end{eqnarray}
Figures 1 and 2 show the dependence of the function
\begin{equation}
<m>=\frac{e^{-\beta(\varepsilon-\mu)}(1+e^{-\beta(\varepsilon q^{-2}
-\mu)})}{1+2e^{-\beta (\varepsilon-\mu)}+
e^{-\beta (\varepsilon(q^{-2}+1)-2\mu)}} \label{m}
\end{equation}
on the energy $\varepsilon$ in comparison with the Fermi function.
\begin{figure}
\centerline{\psfig{figure=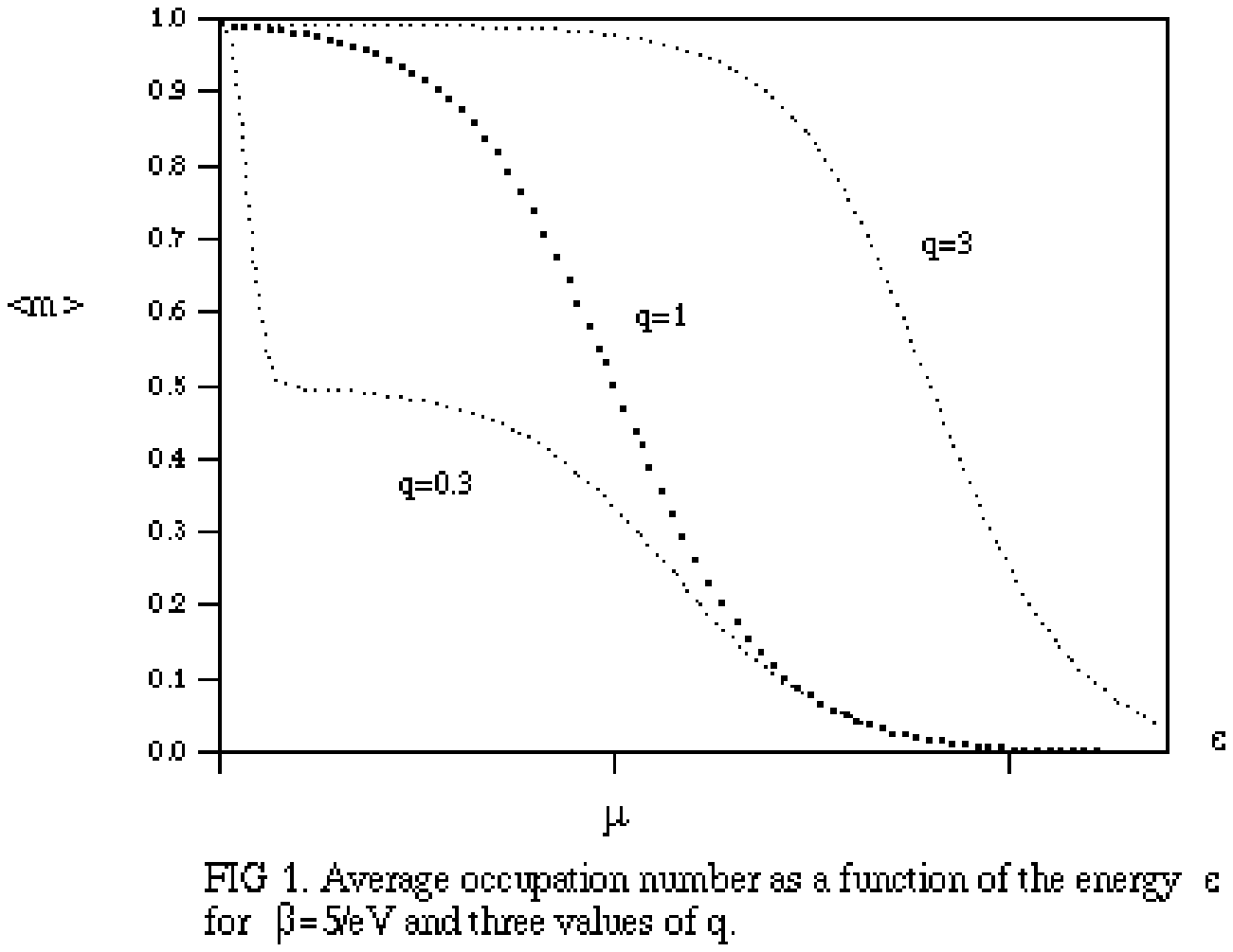,height=4in}}
\end{figure}
Figure 1 shows that for a given temperature 
and larger values of $q$ 
more states become fully occupied. 
For systems with $q<1$  the occupation number $<m>$
remains below the Fermi function for all values of the energy $\varepsilon$.
Figure 2 shows the average occupation number $<m>$ as a function of
$\varepsilon$ for $\beta=200/eV$ as compared with the Fermi function.
For $q<1$, states with energies $\varepsilon$ such that
$q^{2}\mu(0)<\varepsilon<\mu(0)$ have occupation numbers equal to $<m>=1/2$. 
\begin{figure}
\centerline{\psfig{figure=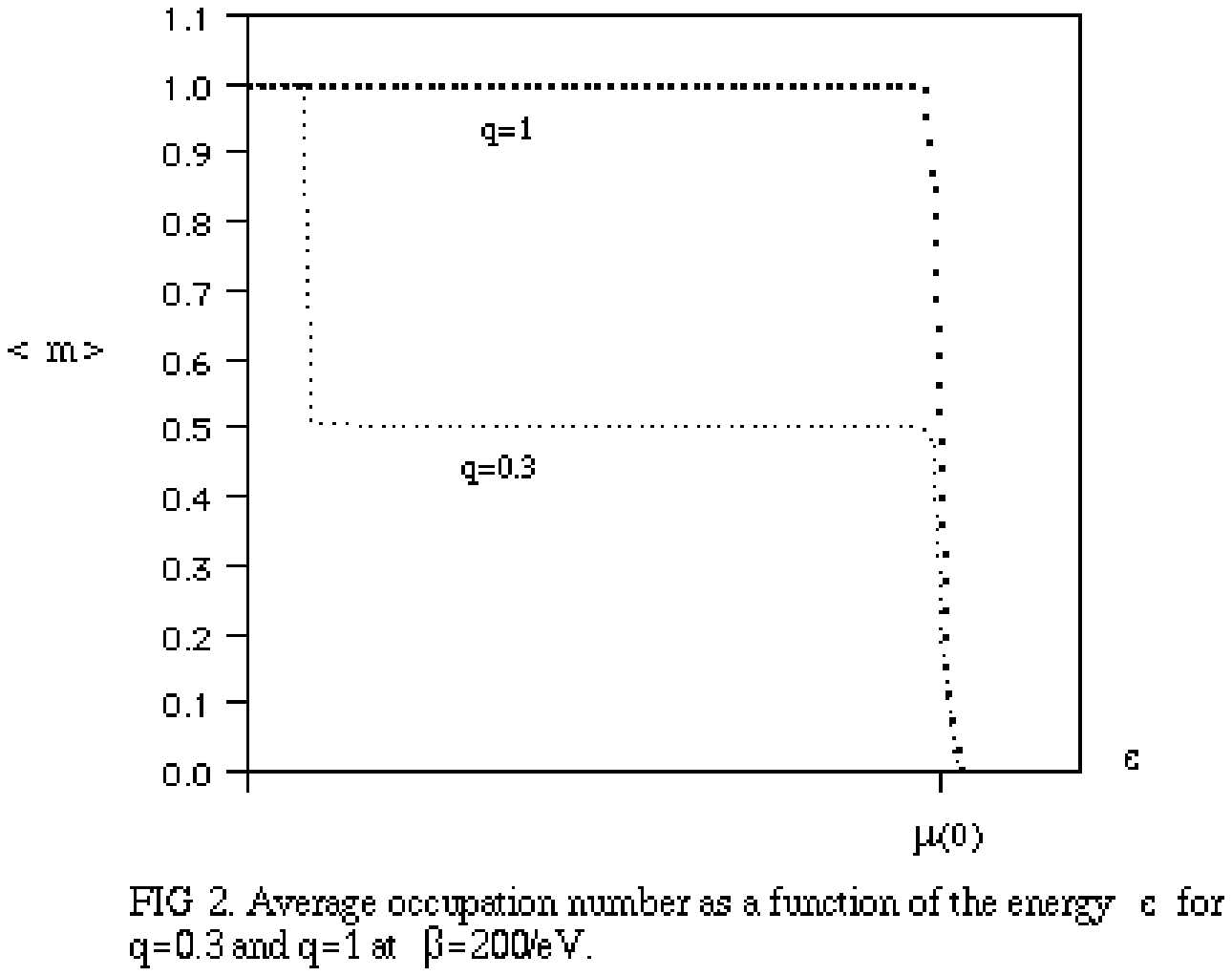,height=4in}}
\end{figure}
The dependence of $<M>$ on the parameter $q$
can be studied by assuming a large volume $V$ 
and particle number such that
we can replace the summations by integrals. Then,  Equation (\ref{M})
becomes 
\begin{equation}
<M>=2V\int_{}^{}\frac{e^{\beta(\mu-\frac{p^{2}}{2m})}
(1+e^{\beta(\mu-\frac{q^{-2}p^{2}}{2m}})d^{3}p}{(2\pi\hbar)^{3}
(1+2e^{-\beta (\frac{p^{2}}{2m}-\mu)}+
e^{-\beta (\frac{p^{2}}{2m}(q^{-2}+1)-2\mu)})}.\label{M1}
\end{equation} 
 In particular,  for $T=0$ and $q<1$ and
the usual definition $\mu(0)=\frac{p(0)^{2}}{2m}$
\begin{eqnarray}
<M>&=&V(2\int_{0}^{qp(0)}\frac{4\pi p^{2}dp}{(2\pi\hbar)^{3}}+
\int_{qp(0)}^{p(0)}\frac{4\pi p^{2}dp}{(2\pi\hbar)^{3}})\nonumber\\
&=&V\frac{(q^{3}+1)}{6\pi^{2}\hbar^{3}}p(0)^{3}\nonumber\\
&=& \frac{(q^{3}+1)}{2}<M_{Fermi}>_{T=0}.\label{Mq<1}
\end{eqnarray}
The internal energy $U$ can be  calculated from
\begin{eqnarray}
U&=&(\frac{\partial\beta\Omega}{\partial\beta}+\mu M)\nonumber\\
&=&V\int_{}^{}\frac{p^{2}}{2m}\frac{e^{\beta(\mu-\frac{p^{2}}{2m})}
(2+(q^{-2}+1)e^{\beta(\mu-\frac{q^{-2}p^{2}}{2m}})d^{3}p}{(2\pi\hbar)^{3}
(1+2e^{-\beta (\frac{p^{2}}{2m}-\mu)}+
e^{-\beta (\frac{p^{2}}{2m}(q^{-2}+1)-2\mu)})}.\label{U}
\end{eqnarray}
At $T=0$ and $q<1$ Equation (\ref{U}) is written
\begin{eqnarray}
U_{q<1}(T=0)&=V&\int_{0}^{qp(0)}(q^{-2}+1)\frac{p^{2}}{2m}\frac{4\pi p^{2}dp}{(2\pi\hbar)^{3}}
+\int_{qp(0)}^{p(0)}\frac{p^{2}}{2m}\frac{4\pi p^{2}dp}{(2\pi\hbar)^{3}}\nonumber\\
&=&\frac{(q^{3}+1)}{2}U_{Fermi}(T=0),
\end{eqnarray}

A simple inspection of Equation (\ref{m}) shows that for $q>1$
and $T=~0$ the function $<m>$ is very similar to the Fermi function.
The only distinction with the Fermi function is that fully occupied states 
are  those
with $\varepsilon<~ \frac{2q^{2}}{q^{2}+1}\mu(0)$. The average number of particles
for $q>1$ becomes then $<M>=q^{3}(\frac{2}{q^{2}+1})^{3/2}<M_{Fermi}>$. 
The internal
energy $U_{q>1}$ at $T=0$ for $q>1$ is similarly calculated to give
$U_{q>1}(T=0)=q^{3}(\frac{2}{q^{2}+1})^{3/2}U_{Fermi}$.  The average energy
per particle $\frac{U}{<M>}$ becomes then independent of $q$ and no
different than the Fermi case.

For low temperatures  we now calculate the chemical potential $\mu$
in the two extreme cases: $q\gg 1$ and $q\ll 1$. 
\begin{description}
\item[a)] $q\gg 1$\\
For very large values of $q$ and $\beta\mu$ Equation (\ref{M1}) can be approximated to read
\begin{equation}
<M>\approx\lambda \int_{0}^{\infty} \varepsilon^{1/2}\frac{d\varepsilon}{1+e^{\beta(\varepsilon-2\mu)}},
\label{q>>1}
\end{equation}
where $\lambda=\frac{V\sqrt{2} m^{3/2}}{\pi^{2}\hbar^{3}}$. If we replace 
the factor $e^{-2\beta\mu}$ by $e^{-\beta\mu}$ in
 Equation (\ref{q>>1})
we obtain the corresponding equation of the Fermi case.  Therefore, for the
case $q\gg 1$
the chemical
potential as a function of the temperature  is not expected to show a  
very different behavior than the standard case.  In fact, Equation (\ref{q>>1})
becomes
\begin{equation}
<M>\approx \frac{2}{3}(2\mu)^{3/2}\lambda+\frac{\pi^{2}\lambda}{12\beta^{2}\sqrt{2\mu}},
\end{equation}
such that after replacement of Equation (\ref{Mq<1}) the chemical
potential  up to terms of
order $(1/\beta)^{2}$ reads
\begin{equation}
\mu\approx \mu(0)(1-\frac{\pi^{2}}{12\sqrt{2}\beta^{2}\mu(0)^{2}}),
\end{equation}
which is identical to the Fermi case except for the factor of
$1/\sqrt{2}$ in the second term.
\item[b)] $q\ll 1$\\
For $q<1$ it is convenient to split Equation (\ref{M1})
into four integrations as follows
\begin{eqnarray}
<M>&=&\lambda\int_{0}^{q^{2}\mu}\varepsilon^{1/2}d\varepsilon-\lambda\int_{0}^{q^{2}\mu}
\varepsilon^{1/2}\frac{1+e^{\beta(\varepsilon-\mu)}}{f(\varepsilon,\mu,q)}d\varepsilon \nonumber\\
&+ &\lambda\int_{q^{2}\mu}^{\mu}
\frac{\varepsilon^{1/2}d\varepsilon}{f(\varepsilon,\mu,q)}+\lambda\int_{\mu}^{\infty}\frac{\varepsilon^{1/2}d\varepsilon}
{f(\varepsilon,\mu,q)},
\end{eqnarray}
where $f(\varepsilon,\mu,q)=e^{\beta(\varepsilon-\mu)}+2+
e^{-\beta(\varepsilon q^{-2}-\mu)}$.
\end{description}

In the limit of $\beta\gg 1$ 
the third integration is simply approximated to 
$I_{3}\approx\int_{q^{2}\mu}^{\mu}\frac {\varepsilon^{1/2}d\varepsilon}
{2}$.  The second and fourth integral require a little more work but
they can be simply solved in the low temperature limit.  For example, the second integral
 becomes approximately equal to
\begin{equation}
I_{2}\approx -\lambda\int_{0}^{q^{2}\mu}\varepsilon^{1/2}\frac{e^{\beta(\varepsilon q^{-2}-\mu)}d\varepsilon}
{2 e^{\beta(\varepsilon q^{-2}-\mu)}+1},
\end{equation}
such that defining a new variable $w=e^{\beta(\varepsilon q^{-2}-\mu)}$ it becomes
\begin{equation}
I_{2}\approx -\frac{\lambda q^{3}}{\beta}\left(\frac{\sqrt{\mu}\ln 3}{2}-\frac{1}
{2\beta\sqrt{\mu}}\int_{1}^{0}\frac{\ln w}{2w+1}dw\right),
\end{equation}
where the first four terms in the expansion of $\ln w$ are sufficient for a good accuracy.
We finally obtain for the average number of particles the result
\begin{equation}
<M>\approx\frac{\lambda\mu^{3/2}(1+q^{3})}{3}+\frac{\lambda \sqrt{\mu}(1-q^{3})\ln 3}{2\beta}
+\frac{ 0.54 \lambda (1+q^{3})}{2\sqrt{\mu}\beta^{2}}.\label{M2}
\end{equation}
In the limit $T=0$ we have that $\mu=\mu_{0}$ and therefore Equation (\ref{M2})
coincides  as expected  with the result in Equation (\ref{Mq<1}).
Once we replace Equation (\ref{Mq<1}) and take $q^{3}\pm 1\approx \pm 1$
we find that for low temperatures the
chemical potential is given by the equation
\begin{equation}
\mu\approx\mu(0)\left(1-\frac{kT\ln 3}{\mu(0)}-\frac{0.24(kT)^{2}}{\mu(0)^{2}}\right),
\end{equation}
which in contrast with the $q=1$ and $q\gg 1$ cases it contains a linear term in $T$.
Figure 3 shows a comparison between the chemical potential for the fermionic case
with the chemical potential of our system for $q\ll 1$ and low temperatures.
\begin{figure}
\centerline{\psfig{figure=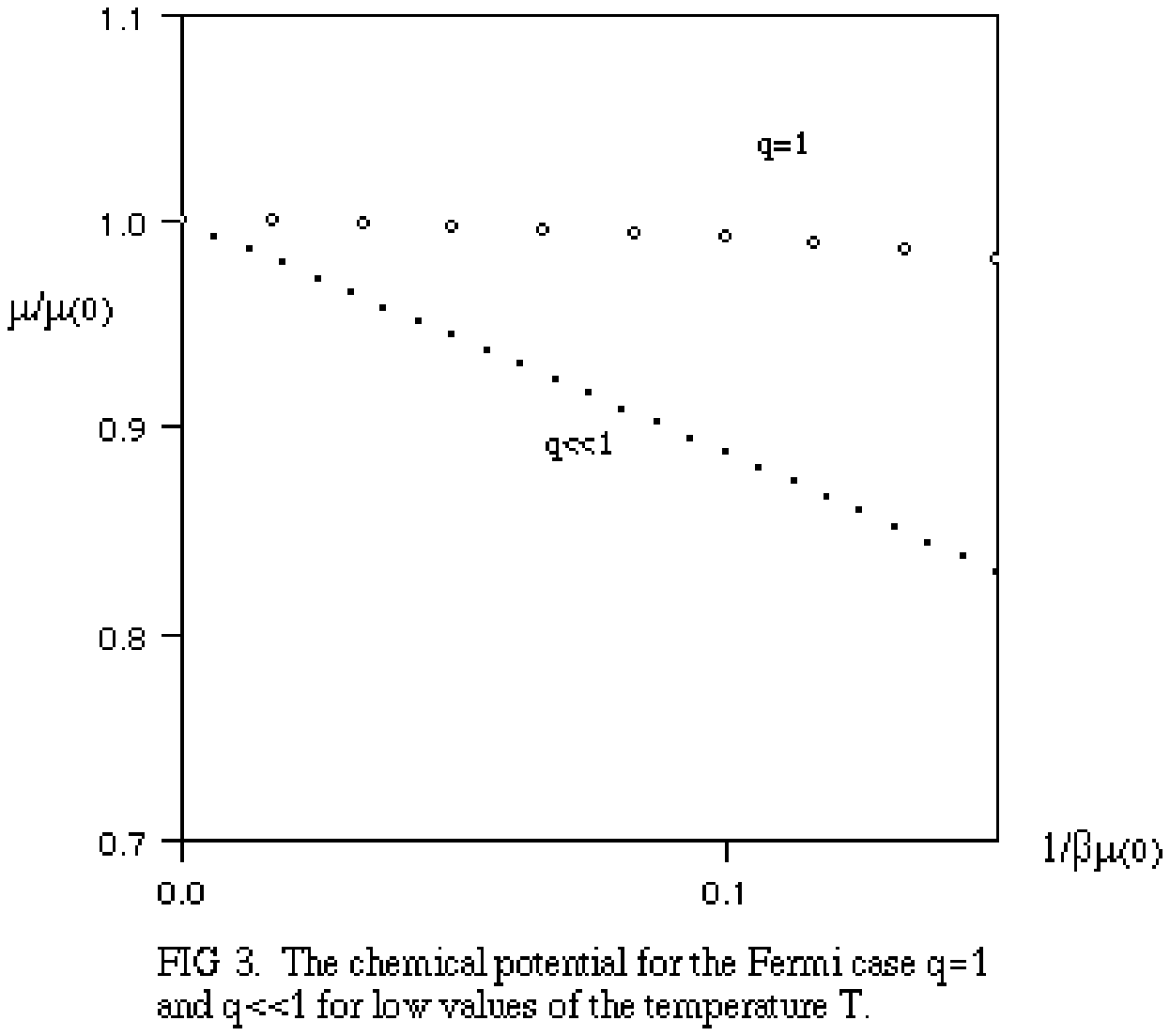,height=3.5in}}
\end{figure}
In conclusion, in this article our main interest was to illustrate 
some of the implications of building
a simple model with quantum group fermionic oscillators.  From the $SU_{q}(N)$ 
fermionic algebra, we first gave a representation of the quantum
group fermionic operators in terms of fermionic fields, and then
considered
the simplest hamiltonian for $N=2$.  This system is equivalent to a theory
with interactions in terms of fermionic number operators, and  clearly
reduces  in the $q=1$ limit to a $SU(2)$ invariant free fermionic system. 
From the grand partition function of the $SU_{q}(2)$ model we obtained
the average occupation number $<m>$ as a function of the energy 
at $T=0$ for $q\neq 1$, and compared it with the
Fermi function.  For $q>1$ and $T=0$ we found that $<m(\varepsilon)>$
is similar to the Fermi function except that the transition point
 from occupied to unoccupied energy states occurs at 
$\varepsilon=\frac{2q^{2}}{q^{2}+1}\mu(0)$
instead of $\varepsilon=\mu(0)$. For $q<1$ the function  $<m(\varepsilon)>$
goes from $<m>=1$ to $<m>=1/2$ at $\varepsilon=q^{2}\mu(0)$
and from $<m>=1/2$ to $<m>=0$ at $\varepsilon=\mu(0)$,
departing then considerably from the standard case.  A calculation of the
chemical potential for $q\gg 1$ and low $T$ results in  no interesting new behavior 
as compared 
with the $q=1$ case. For $q\ll 1$, quantum group symmetries
have an effect on the chemical
potential, as shown in Figure 3, through a linear temperature dependent term.

Therefore, as far as
statistical physics is concerned,  we should expect that
new interesting  consequences of introducing quantum group symmetries
in a fermionic model will be found within the values  $0<q<1$ rather than for $q>1$.
Whether the same
paradigm applies to the case of a quantum group bosonic system is a 
question we hope to address in a separate publication.

\end{document}